\RequirePackage{fix-cm}
\documentclass[onecolumn,epjc3,longbibliography]{svjour3}  

\smartqed  
\RequirePackage{graphicx}
\journalname{Eur. Phys. J. C}

\usepackage{geometry}
\usepackage{xcolor}
\usepackage{amssymb}
\usepackage{cite}
\usepackage{amsmath}
\usepackage[normalem]{ulem}
\usepackage{braket}
\usepackage{nccmath}

\newcommand{\bal}{\begin{align}}
\newcommand{\eal}{\begin{align}}

\def\cA{{\cal A}}
\def\cU{{\cal U}}
\def\cP{{\cal P}}

\def\cT{{\cal T}}
\def\cU{{\cal U}}

\def\cH{{\cal H}}

\def\cW{{\cal W}}

\newcommand{\secn}[1]{Section~1}
\newcommand{\appn}[1]{Appendix~1}

\long\def\comment#1{ }

\def\Tr{\text{Tr}}

\def\ua{\uparrow}
\def\da{\downarrow}

\def\and{\quad\text{and}\quad}

\newcommand{\rme}{{\rm e}}

\def\q{{\boldsymbol q}}
\def\0{{\boldsymbol 0}}
\def\p{{\boldsymbol p}}

\def\k{{\boldsymbol k}}
\def\m{{\boldsymbol m}}
\def\n{{\boldsymbol n}}
\def\x{{\boldsymbol x}}
\def\y{{\boldsymbol y}}
\def\X{{\boldsymbol X}}

\def\0{{\boldsymbol 0}}

\def\P{{\boldsymbol P}}
\def\X{{\boldsymbol X}}

\begin{document}

\title{A quantum strategy to compute the jet quenching parameter $\hat{q}$ 
}


\author{ 
        Jo\~{a}o Barata\thanksref{e2,addr1}
        \and
        Carlos A. Salgado\thanksref{e3,addr1}
}

\thankstext{e2}{joaolourenco.henriques@usc.es}
\thankstext{e3}{carlos.salgado@usc.es}


\institute{
           Instituto Galego de F\'{i}sica de Altas Enerx\'{i}as (IGFAE), Universidade de Santiago de Compostela,
E-15782 Galicia, Spain \label{addr1}
}

\date{Received: date / Accepted: date}

\maketitle

\begin{abstract}
Jet quenching, the modification of the properties of a QCD jet when the parton cascade takes place inside a medium, is an intrinsically quantum process, where color coherence effects play an essential role. Despite a very significant progress in the last years, the simulation of a full quantum medium induced cascade  remains inaccessible to classical Monte Carlo parton showers. In this situation, alternative formulations are worth being tried and the fast developments in quantum computing provide a very promising direction. The goal of this paper is to introduce a strategy to quantum simulate single particle momentum broadening, the simplest building block of jet quenching. Momentum broadening is the modification of the quark or gluon transverse momentum due interactions with the underlying medium, modeled as a QCD background field. At the lowest order in $\alpha_s$ that we consider here, momentum broadening does not involve parton splittings and particle number is conserved, greatly simplifying the quantum algorithmic implementation. This quantity is, however, very relevant for the phenomenology of RHIC, LHC or the future EIC. 



\end{abstract}

\section{Introduction}
\label{sec:intro}

The idea of simulating the dynamics of complex quantum physical systems by using other simpler and controllable quantum systems (which we shall refer to as quantum computers) was first realized by Feynman in the 80's~\cite{Feynman:1981tf}. Since then quantum simulation as seen widespread application in physics, chemistry and many other areas~\cite{Georgescu:2013oza,Schuld_2014,Or_s_2019}.

In recent years, a big effort has been made towards exploring to what extent quantum computers might enhance our understanding of High Energy/Nuclear Physics (HEP/NP) phenomena~\cite{carlson2018quantum,cloet2019opportunities,matchev2020quantum}. Some of the most recent studies in the application of quantum computing to HEP/NP phenomenology have resulted in the formulation of quantum parton showers~\cite{Bepari:2020xqi,Bauer:2019qxa}, quantum jet clustering algorithms~\cite{Pires:2021fka,Wei:2019rqy}, digital simulation of effective field theories~\cite{Bauer:2021gup} and of the propagation of hard probes in a thermal QCD bath~\cite{deJong:2020tvx}.

In this paper, we propose a strategy to use a quantum digital computer to simulate the evolution of a single parton in the presence a QCD background field. In particular, we are interested in the $\alpha_s^0$ effect, corresponding to the broadening of the parton's momentum. Although this effect has been extensively studied in jet quenching theory~\cite{Gyulassy:2002yv,Blaizot:2012fh}, it is only easily computed for isotropic and homogeneous media, where the field fluctuations behave as \textit{white noise}. 
More interestingly, at the amplitude level, the associated in-medium propagators are the building blocks of jet quenching formulation for e.g. medium-induced gluon radiation. For this reason, we argue that our algorithmic implementation can be considered as a first step towards a complete simulation of the in-medium parton cascade with quantum color coherence.

We consider an energetic parton that emerges from a hard scattering event and then propagates inside a QCD medium. The net effect of the medium is to alter the initial transverse momentum of the parton. The underlying gauge field is treated stochastically, in line with the usual approach employed in jet quenching theory and phenomenology. As a consequence, our algorithm consists in a hybrid classical-quantum strategy~\cite{mueller2020deeply,PhysRevX.6.031045,deJong:2020tvx}, with the parton evolution in time being tracked, at the amplitude level, by the quantum computer and the gauge fields being provided as an input to the circuit. 
We will not make an attempt to improve here on these dual description as an eventual future implementation of the quantum computation of the gauge fields would be straightforward to implement in our procedure.

Although for an actual implementation, crucial aspects such as quantum error correction~\cite{Terhal_2015}, encoding details or Trotter error analysis~\cite{Childs_2021} have to be considered, we will mostly stay at a more conceptual level and leave such an analysis for future work. In addition, we will try to highlight the connection between our approach and \textit{standard} jet quenching treatment of momentum broadening, thus making what follows more relevant for an interested reader.

The present manuscript is divided as follows: section~\ref{sec:HP_in_A} briefly reviews the physics of a hard parton propagating in a gauge classical background field, while in section~\ref{sec:QS_for _P} we provide the equivalent Hamiltonian formulation. In the remainder of the section we detail how such a problem can be implemented in a digital quantum computer. Finally, in section~\ref{sec:non_abelian} we detail how to deal with non-trivial evolution in color space and finally section~\ref{sec:Conclusion} gives the paper's main conclusions and outlines possible future research paths. Details on most sections are provided in four appendices.

\section{Hard parton propagating in a background field}
\label{sec:HP_in_A}
In this section we review the theoretical description of a highly boosted parton propagating in a classical background gluon field $\cA$. We will consider the cases where the propagating parton, a quark, is in the singlet or fundamental color representations, though we will mostly assume that the evolution in color space is trivial. Detailed and ample discussions on jet quenching theory can be found, for example, in~\cite{Blaizot:2012fh,CasalderreySolana:2007zz,Blaizot:2015lma, Mehtar-Tani:2013pia} and references therein.

We begin by considering a highly energetic quark, being produced inside a QCD medium (whose exact origin is not relevant) from a hard process. Due to the highly boosted kinematics, the quark's four-momentum $p=(p^0,\p,p^z)$ is more conveniently expressed in light-cone coordinates $p\equiv(\omega,\p,p^-)=((p^0+p^z)/2,\p,p^0-p^z)$, with $\p$ the transverse momentum, $\omega$ the light-cone energy and $p^-$ the minus component of the quark's momentum. The quark is assumed to be moving in the plus direction, so that $\omega$ is the large momentum component. 

It is also convenient to work in the light-cone gauge for the background field $\cA^\mu$, taking $\cA^+=0$ and fixing the residual gauge freedom such that $\cA^-$ is the only non-vanishing component of the background field~\cite{Blaizot:2008yb}. Additionally, because $p^+$ is the large component of the quark's momentum, its interaction with the gluon field is highly localized in $x^-$, so that we can simplify the spacetime dependence of the field to be $\cA^-(x^+,\x,x^-)\equiv \cA(x^+,\x,0)$,  dropping the $x^-$ dependence in what follows.

Finally, in the boosted regime the local quark-field spin-flip interactions are energy suppressed and can be ignored. Thus, for each field insertion $\cA^-$ and in the strict high energy limit, the large momentum component $\omega$ and transverse component $\p$ are conserved, with the quark state acquiring an eikonal phase. It is however usual to relax this approximation and allow for a small transverse momentum transfer at each vertex, while light-cone energy is still conserved. Accounting for this sub-eikonal corrections, the quark propagation in the QCD field can be reduced to the study of a two dimensional non-relativistic quantum system~\cite{Blaizot:2015lma}.

To make this discussion more quantitative, we consider the in-medium scalar quark propagator $G(t,\x;0,\y)$ in the transverse plane, between spacetime points $(0,\y)$ and $(t,\x)$~\cite{Blaizot:2012fh}. This propagator is the Green's function to the following two dimensional Schrodinger equation
\begin{equation}\label{eq:Sch_G}
\left(i\partial_t+\frac{\partial^2_\x}{2\omega}+g\cA^-(t,\x)\cdot T\right)G(t,\x;0,\y)=i\delta(t)\delta(\x-\y) \, ,
\end{equation}
where we have contracted the background gauge field with the respective $SU(3)$ generators $T$ in the adequate representation. The remaining terms are diagonal in color space.
This equation explicitly shows that the quark propagation is equivalent to a non-relativistic two dimensional quantum mechanical system, describing a single particle evolving in time with the Hamiltonian~\cite{Blaizot:2015lma}
\begin{equation}\label{eq:Hamiltonian}
\cH(t)=\frac{\p^2}{2\omega}+g\cA^-(t,\x)\cdot T=\cH_K + \cH_\cA(t)   \, , 
\end{equation}
where $\omega$ plays the role of a mass and light-cone time plays the role of time\footnote{In the boosted regime, light-cone time $x^+$ becomes the same as time $x^0$ since $x^+=(x^z+x^0)/2\sim x^0$.}. In the strict eikonal limit, where $\frac{\p^2}{\omega}\to 0$, the kinetic term drops out and the evolution leads to the state acquiring a field dependent phase, as mentioned above. 

\section{Quantum simulating momentum broadening}\label{sec:QS_for _P}
From Eq. \ref{eq:Hamiltonian} one can construct the time evolution operator (with $\cT$ the time ordering operator)
\begin{equation}\label{eq:U_full}
\cU(t,0)  \equiv   \cT \exp\left[-i\int_0^t ds\, \cH(s)\right] \, , 
\end{equation}
which acts on the infinite dimensional Hilbert space of a single free particle in two spatial dimensions, such that from an initial state $\ket{\psi_0}$ at time $t=0$ one obtains the time evolved state $|\psi_t\rangle $ via
\begin{equation}\label{eq: opp}
|\psi_t\rangle = \cU(t,0)|\psi_0\rangle  \, .
\end{equation}
The Hilbert space is spanned by the position eigenvectors $\ket{\x}$ or by their Fourier pair $\ket{\p}$. These two bases are convenient since $\hat{\p}\ket{\p}=\p\ket{\p}$ and $\hat{\cA}^{-a}(t,\hat{\x})\ket{\x}=\cA^{-a}(t,\x)\ket{\x}$, where we used the hats to highlight the difference between operators and c-numbers; we also used the fact that the quark-medium interaction is localized in position space (and conversely delocalized in momentum space). 

\begin{figure}[h!]
    \centering
    \includegraphics[width=0.8\textwidth]{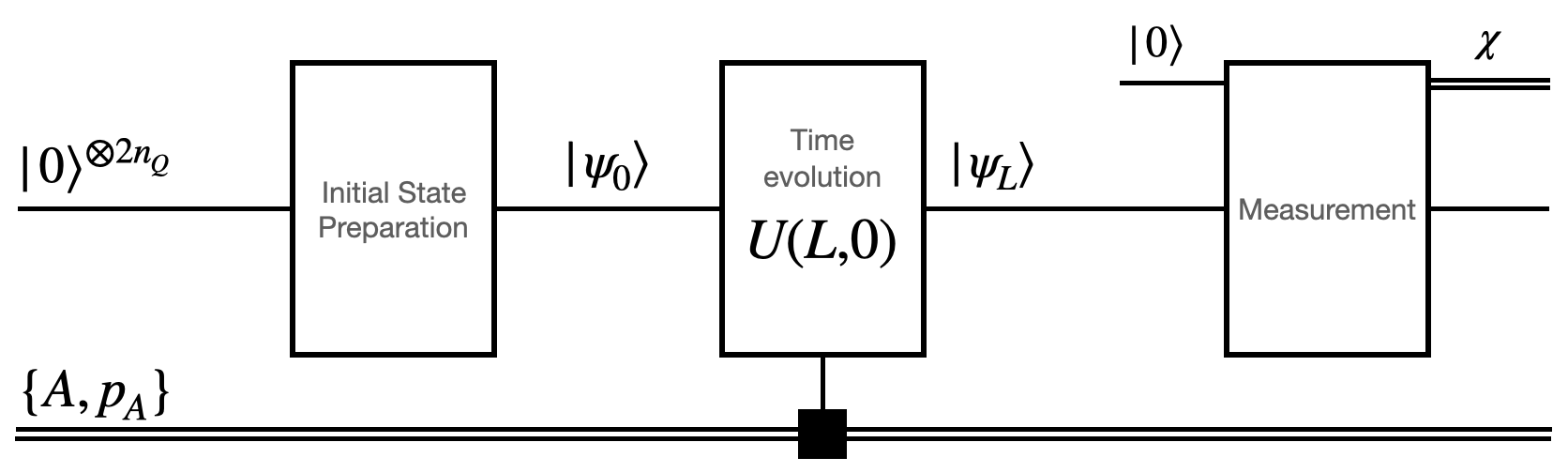}
    \caption{Overview of the quantum circuit detailed in the main text. Single lines denote quantum channels while double lines denote classical ones. Above each line we detail the state being store in the circuit (see main text for notation). The $\blacksquare$ denotes that the time evolution gates parameters are to be determined from the field $A$.}
    \label{fig:1}
\end{figure}

We now detail how to frame single particle momentum broadening in terms of a digital quantum simulation algorithm, implementing Eq.~\ref{eq: opp}. The algorithm, summarized in Fig. \ref{fig:1}, can be divided as follows:

\begin{enumerate}
    \item \textbf{Input} -- i) Template distribution to be loaded as an initial state $\ket{\psi_0}$ ii) A list of $m$ field configurations $\cA^-$ with the associated weights $p_{\cA^-}$, storing the probability of generating each configuration;
    
    \item \textbf{Encoding} -- Map between the degrees of freedom of the quantum system and the qubits;
    
    \item \textbf{Initial state preparation} -- Preparation of $\ket{\psi_0}$;
    
    \item \textbf{Time evolution} -- Implementation of Eq.~\ref{eq: opp};
    
    \item \textbf{Measurement} -- Retrieving physical information by measuring the qubits, according to a sensible protocol;
    
    \item \textbf{Output} -- For each field configuration the algorithm will output the expected value of a random variable $\chi$, which should be then medium-averaged over all $m$ configurations.
\end{enumerate}

\subsection{Encoding}\label{sec:enconding}
We begin by discretizing the problem in position space, such that $\ket{\x}=\ket{a_s\n }$, with $a_s$ the spatial lattice spacing and $\n=(n_1,n_2)$ a two component dimensionless transverse vector, where each component can take integer values between $0$ and $N_s-1$, with $N_s$ the number of lattice sites per dimension. The spatial cutoff is given by $\x_{\rm max}=a_s(N_s-1,N_s-1)$. Also, the spatial discretization induces a lattice discretization in momentum space with $\ket{\p}=\ket{a_d \q}$ and $a_d=\frac{2\pi}{a_sN_s}$ the momentum space lattice spacing, with $\q=(q_1,q_2)$ a two dimensional vector with each component also taking integer values between $0$ and $N_s-1$\footnote{In the following discussion we will consider only positive values for the position and momentum of the quark, see~\ref{app:discretization}.}. 

One can rewrite the Hamiltonian $\cH$ in terms of the dimensionless Hamiltonian $H=\cH a_s$ (see~\ref{app:discretization})
\begin{equation}\label{eq:dimless_H}
H=\frac{\P^2}{2 E} + g A(t,\X)\cdot T = H_K + H_A(t)   \, , 
\end{equation}
with $\hat{\P}\ket{\q}=\q\ket{\q}$ and $\hat{\X}\ket{\n}=\n\ket{\n}$ the dimensionless position and momentum operators. Also $A(t,\n)\cdot T=a_s \cA^-(t,a_s\n )\cdot T$ and $E=\frac{N_s^2 \omega a_s}{4\pi^2}$ is the dimensionless energy factor. In what follows, position and momentum vectors are assumed to be given in this dimensionless basis.

With this discretization, the problem can be mapped to the qubits available in a quantum computer. For each spatial dimension, we use a register with $n_Q$ qubits (each qubit being equivalent to a $1/2$-spin), such that we can generate $2^{n_Q}=N_s$ states. We use the QIS convention~\cite{NielsenChuang} to denote the single up spin state $\ket{\ua}=\ket{0}=[1,0]^T$ in the computational basis (with the last equality giving the associated vector representation) and $\ket{\da}=\ket{1}=[0,1]^T$. Then, any component of the vector $\ket{\n}$ can be represented by a product of many spins, in a binary basis (see \ref{app:discretization}). The associated momentum state vector $\ket{\q}$ 
is obtained by applying a standard quantum Fourier Transform  (qFT).

\subsection{Initial state preparation}\label{sec:ISTP}
Given the above encoding, the first step in the algorithm consists in loading a desired template distribution by constructing the initial state $\ket{\psi_0}$ from the fiducial state $\ket{0}^{\otimes 2n_Q}$. The template is meant to represent the relevant physics of the hard scattering which generates the initial parton.

 In this manuscript, since we are interested in extracting the jet quenching parameter $\hat{q}$ from the quantum simulation output, we wish to avoid contributions coming from initial state physics.  Therefore, we shall mainly focus on the case where $\ket{\psi_0}=\ket{\p=\0}$. 

However, including a localized initial state distribution might be important for certain digitizations where one can not prepare the state $\ket{\p=\0}$ exactly or if one is simply interested in studying how different production mechanisms influence the final state. Several strategies to prepare $\ket{\psi_0}$ from an integrable template distributions can be found in the literature~\cite{Barata:2020jtq,kitaev2009wavefunction,grover2002creating,kaye2004quantum}. Depending exactly on what $\ket{\psi_0}$ one wants to prepare, in principle, one can devise a routine which only requires $\mathcal{O}(n_Q)$ basic quantum gate operations. 

\subsection{Time evolution}\label{sec:Time_evolution}
After the initial state $\ket{\psi_0}$ has been prepared, we time evolve it for a time $L$, producing the final state $\ket{\psi_L}$. The time evolution operator in Eq.~\ref{eq:U_full} can be written in terms of the dimensionless Hamiltonian $H$ and medium length $L^\prime\equiv L/a_s$\footnote{In general, one could choose another length scale to make time dimensionless, leading to the appearance of a ratio between $a_s$ and such scale.}. 
\begin{equation}\label{eq:U}
U(L^\prime,0)  \equiv   \cT \exp\left[-i\int_0^{L^\prime} dt\, H(t)\right] \, .    
\end{equation}
Directly implementing $U(L^\prime,0)$ in terms of a quantum circuit is in general impossible. Rather, one decomposes the full evolution into a sequence of short time evolution steps. Here we do this by considering the simplest product formula~\cite{Poulin_2011}, decomposing $U$ as
\begin{ceqn}
\begin{equation}\label{eq:U_TS}
U(L^\prime,0)\approx   \prod_{k_t=1}^{N_t}    \left\{ \exp\left[-i H_K \frac{L^\prime}{N_t}\right]\exp\left[-i H_A\left(k_t\cdot \frac{L^\prime}{N_t}\right) \frac{L^\prime}{N_t}\right]\right\} \equiv \prod_{k_t=1}^{N_t}    \left\{ U_K(\varepsilon_t) U_A(k_t\cdot \varepsilon_t,\varepsilon_t)\right\} \, ,
\end{equation}
\end{ceqn}
where we have effectively sliced time into $N_t$ steps, each with a length $\varepsilon_t\equiv L^\prime/N_t$. In each time step, the evolution operator is split into a short evolution according to $H_K$, followed by an evolution in time with $H_A$. Notice that during the time interval $(k_t\cdot \varepsilon_t,(k_t+1)\cdot \varepsilon_t)$ the field $A$ is taken to be constant, leading to the constraint $\varepsilon_t^{-1}\gg ||\partial_t H_A(t)||$; there exist algorithms~\cite{Poulin_2011} which circumvent this constraint, as well as other strategies (see for example~\cite{Berry_2015,Berry_2020,Wiebe_2010,Berry_2014}) to quantum simulate time dependent Hamiltonians with expected higher precision. Although the way one chooses to implement $U$ is of critical importance to determine the efficiency and accuracy of the quantum circuit, since we are aiming to restrict our discussion to a more conceptual level, we limit our analysis to the simple product formula considered above, which has a Trotter error $\mathcal{O}(\varepsilon_t^2)$.

\begin{figure}[h!]
    \centering
    \includegraphics[width=0.8\textwidth]{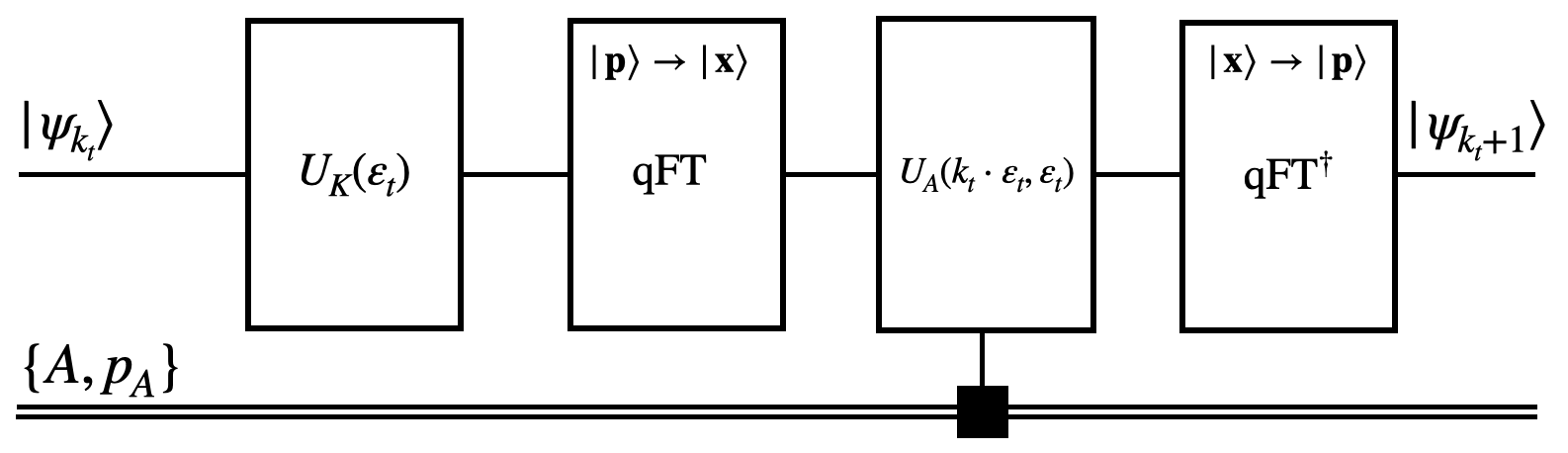}
    \caption{Outline of the implementation of the time evolution operator $U$. Here we detail the $k_t^{\rm th}$ time step, as indicated in Eq.~\ref{eq:U_TS}.}
    \label{fig:2}
\end{figure}

Let us now consider the $k_t^{\rm th}$ time slice of the evolution. As mentioned above, $H_K$ has a trivial action in the momentum basis, while $H_A$ can be simply written in the position basis; this justifies the decomposition of $H$ taken in Eq.~\ref{eq:U_TS}. Since these bases are trivially related by a qFT, one can simply first time evolve with $H_K$, perform the transformation $\ket{\p}\to \ket{\x}$, time evolve with $H_A$ and transform back to the $\ket{\p}$ basis, the generated state being the input to the $k_{t}+1^{\rm th}$ time slice; this strategy is illustrated in Fig.~\ref{fig:2}. 

The time evolution operator $U_K$ is diagonal in the $\ket{\p}$ basis
\begin{equation}
\begin{split}
U_K(\varepsilon_t) \ket{\p}= \exp\left(-i\frac{\varepsilon_t}{2E}\p^2\right) \ket{\p} \, ,    
\end{split}    
\end{equation}
thus one only needs to implement a circuit which generates a state dependent phase. This can be achieved using the algorithm introduced in~\cite{Zalka_1998}, which we detail in~\ref{app:Zalka}. 

After performing the qFT, using standard implementations of the circuit~\cite{NielsenChuang}, one has to compute the action of $U_A$. Although this operator is diagonal in the $\ket{\x}$ basis,
\begin{equation}\label{eq:UA}
\begin{split}
U_A(k_t\cdot \varepsilon_t,\varepsilon_t) \ket{\x}= \exp(-ig\varepsilon_t A(k_t\cdot \varepsilon_t,\x)) \ket{\x} \, ,    
\end{split}    
\end{equation}
the value of the phase depends on the local value of the field $A$. Notice that here we assume that the quark is a color singlet; see section \ref{sec:non_abelian} for details on how to deal with non-trivial color evolution. In principle, one could again use a strategy similar to the one used to implement $U_K$ (see \ref{app:Zalka}). However, this assumes that one could construct $N_t$ oracles which quantum compute $A(k_t\cdot \varepsilon_t,\x)$ for every $\x$ in each time slice. Since in general one does not have a closed form expression or a simple numerical routine to compute the field values, such an approach might not be possible.

A more feasible approach would consist on first computing the field values for all positions and times. This would require evaluating the field $\mathcal{O}(N_t\times N_s^2)$ times, which would defeat the purposes of the present strategy since it requires exponentially many classical evaluations of $A$. Nonetheless, we notice that in practice a small number of qubits $n_Q$ is needed to have a sufficiently good discretization (see section~\ref{sec:numerical_estimates}), and thus the actual number of field evaluations needed could in practice be performed by a classical computer.

Once one has evaluated all the relevant field values, they are stored in a classical memory (double lines in Figs. \ref{fig:1} and \ref{fig:2}) which are loaded onto the circuit as parameters to the basic gates implementing Eq.~\ref{eq:UA}. We illustrate this procedure in~\ref{app:Zalka}, that requires solving a system of linear equations with $N_s^2$ independent variables, which following the same arguments as above should be doable in practice, at least for near term small system applications\footnote{We note however that this linear system only has to be solved once for each $n_Q$.}. Clearly the implementation of the operator $U_A$ would greatly benefit from native implementations of quantum diagonal gates, where each entry exponentiates a circuit input~\cite{Heeres_2015}\footnote{Quantum strategies to simulate the time evolution of the background field could also be coupled to our strategy. This could in principle simplify the implementation of $U_A$.}.

After performing this operation and transforming back to the momentum basis, this block is iterated until $k_t=N_t$, where the time evolution section of the algorithm terminates.

\subsection{Measurement}\label{sec:Measurement}
Having prepared the state $\ket{\psi_L}=\sum_\q \psi_L^\q \ket{\q}$ one could simply measure all the $2n_Q$ qubits, obtain the probabilities $|\psi_L^\q|^2$ for every $\q$ and reconstruct the underlying probability distribution. However, such a strategy requires a exponentially large number of measurements. This constraint is a direct consequence of the quantum nature of the simulation, absent from classical simulations where information can be easily retrieved.

In this section, we assume that the initial condition of the state was that of a quark with $\p=\0$. In this case the coefficients $|\psi_L^\q|^2$ are directly related to the single particle broadening distribution; see \ref{app:broad}. This statement is only true after having averaged over all field configurations, the so called medium average, which in our strategy is performed at the end of the algorithm. For each of the $m$ field configurations one runs the algorithm the necessary number of times to extract the expectation value of some classical variable  $\chi$ (to be detailed below). Then, one averages over all $m$ expectation values
\begin{equation}
\langle \chi \rangle_{\rm M} = \frac{1}{\sum_{i=1}^{m} p_{A^{(i)}}} \sum_{i=1}^{m} p_{A^{(i)}} \langle \chi \rangle_{\rm QM}^{(i)}  \, ,
\end{equation}
where $p_{\cA^-}=p_{A}$, the $i$ superscript denotes a particular field configuration, running up to $m$, and $\langle . \rangle_{\rm M}$ denotes the average over field configurations while $\langle . \rangle_{\rm QM}$ denotes the (quantum mechanical) expectation value. 

The numerical value for $m$ depends on field fluctuations. In jet quenching, these are typically Gaussianly distributed, following the prescription of the Mclerran-Venugopalan (MV) model~\cite{McLerran:1993ka,McLerran:1993ni} and are encapsulated in the field-field correlator~\cite{Kovchegov:1996ty,Kovchegov:1997pc,McLerran:1993ka,McLerran:1993ni}. We note however that in our approach, one is not constrained to assume the MV prescription, nor does one need to explicitly construct any field correlator. In addition, due to the formal similarities between jet quenching and saturation physics~\cite{MehtarTani:2006xq}, the physical origin of $\cA^-$, either generated from hot and dense quark gluon plasma, the initial glasma or from cold nuclear matter, is not constrained. This means that our approach should be able to explore the evolution of the jet quenching parameter $\hat{q}$, both in time and in orthogonal spatial directions~\cite{Ipp:2020nfu}, for different medium models. The only practical constraint is that the larger the background field fluctuations become, the larger $m$ must be, despite this only leading to a linear increase in cost for running the full algorithm.

\begin{figure}[h!]
    \centering
    \includegraphics[width=0.8\textwidth]{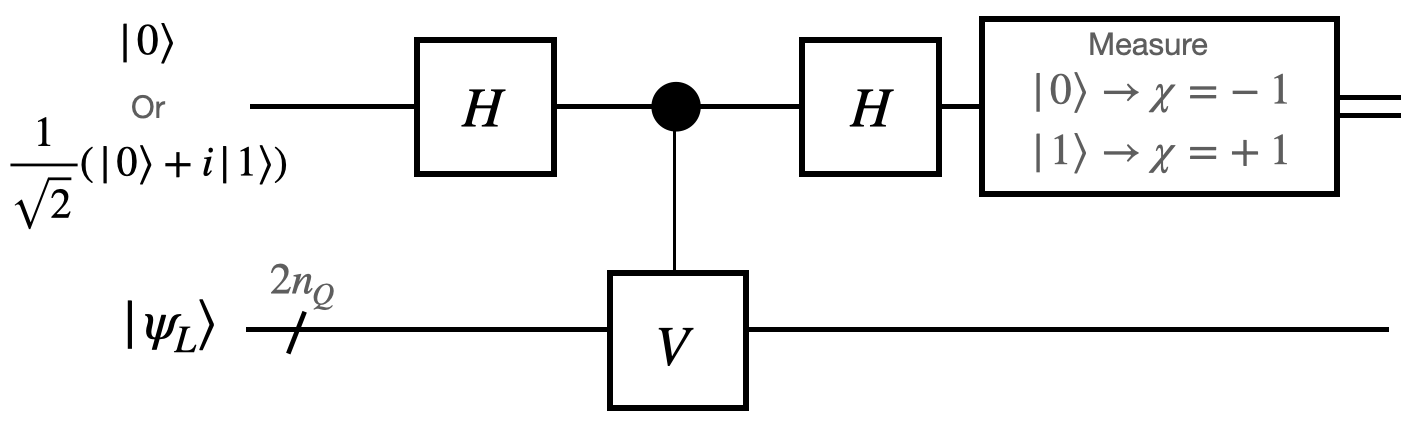}
    \caption{Circuit representation of the measurement strategy.}
    \label{fig:meas}
\end{figure}

We then focus the remaining discussion on the case of a fixed field configuration and how to extract $\hat{q}$ for that $\cA^-$. We add an ancilla qubit to the circuit and perform the Hadamard test detailed in Fig.~\ref{fig:meas}.

One first transforms the ancilla, which can be either prepared in the state $\ket{0}$ or in the superposition $1/\sqrt{2}(\ket{0}+i\ket{1})$, by the Hadamard gate $H=H^\dagger$, and then applies a unitary transformation $V$ on the physical state if the ancilla is in the state $\ket{1}$. Finally the transformation on the ancilla is reversed and one measures the qubit. We associate the measured value to a random variable $\chi$ which takes the values $-1$ if we observe the state $\ket{0}$ and $+1$ if the state $\ket{1}$ is generated. This strategy is not the only possible one, but it is particularly simple and inexpensive in terms of extra ancillas and number of gate operations.

One can show that if the ancilla is in the initial state $\ket{0}$ (see \ref{app:meas}), then 
\begin{equation}
\langle \chi \rangle_{\rm QM}\equiv\bra{\psi_L} V+V^\dagger \ket{\psi_L}= \Re    \bra{\psi_L} V \ket{\psi_L} \, . 
\end{equation}
On the other hand, if the ancilla is prepared in the state $1/\sqrt{2}(\ket{0}+i\ket{1})$, we have that 
\begin{equation}
\langle \chi \rangle_{\rm QM}= \Im    \bra{\psi_L} V \ket{\psi_L} \, ,
\end{equation}
which when combined give access to both the real and imaginary parts of the expectation value of the unitary operator $V$.

Let us consider first the case where $V=V_\alpha=\exp(i\alpha \P^2)$. Then
\begin{equation}
\Re    \bra{\psi_L} V_\alpha \ket{\psi_L} =\langle \cos (\alpha \P^2) \rangle_{\rm QM} \, ,
\end{equation}
and 
\begin{equation}
\Im   \bra{\psi_L} V_\alpha \ket{\psi_L} = \langle \sin (\alpha \P^2) \rangle_{\rm QM}   \,,
\end{equation}
from which one extracts $\langle e^{i\alpha \P^2}\rangle_{\rm QM}$, by definition. We also have that 
\begin{equation}\label{eq:generating_func}
\begin{split}
\langle e^{i\alpha \P^2}\rangle_{\rm QM}&=1+\sum_{k=1}^\infty \frac{i\alpha^k}{k!}\langle \langle  2k\rangle \rangle \, , 
\end{split}
\end{equation}
where $\langle \langle  2k\rangle \rangle \equiv \langle \P^{2 k}\rangle_{\rm QM}$ corresponds to the expectation value of the $2k$ power of the momentum operator. Eq.~\ref{eq:generating_func} can be viewed as the (even) moment generating function, and it is easily related to the cumulants of the underlying broadening distribution. Also, in the case where initial state effects are absent, $a_d^2\langle \langle  2\rangle \rangle =\hat{q}L$, where we inserted $a_d^2$ to get the correct dimensions.

Furthermore, one has the freedom to vary $\alpha$ such that, for small enough  $\alpha$, only linear variations are relevant
\begin{equation}
\langle e^{i\alpha \P^2}\rangle_{\rm QM}\approx 1 +i \frac{\alpha}{a_d^2} \hat{q}L\to \langle \sin(\alpha \P^2)\rangle_{\rm QM} \approx \frac{\alpha}{a_d^2} \hat{q}L \, .
\end{equation}
Notice that the left hand side corresponds to a quantity readily extracted from the quantum computer, while the right hand side is written in terms of the physical jet quenching parameter. 

If one includes higher order $\alpha$ corrections, then one has access to the even moments of the momentum distribution and the respective cumulants. One can thus imagine varying $\alpha$ and from the observed evolution retrieving $\langle \langle  2k\rangle \rangle $ moments via a numerical fit. Of course, such a strategy, on top of the additional polynomial cost in $m$, would increase the cost of running the algorithm by the number of $\alpha$ values to be explored.

If one is only interested in extracting $\hat{q}$ (which is the most relevant medium parameter for jet quenching), one could consider the unitary $V=\exp(iF(\P^2))$, with $F(\P^2)=\arccos(\P^2)$. Then, for the case where the ancilla is initially set to $\ket{0}$, we obtain
\begin{equation}
\langle X\rangle_{\rm QM}=    \bra{\psi_L} \cos(\arccos(\P^2)) \ket{\psi_L}= \langle \langle  2\rangle \rangle \, .
\end{equation}
In principle, one could implement this protocol following~\cite{Zalka_1998} (see also~\ref{app:Zalka}), provided an efficient arithmetic oracle could be constructed.

\section{Treating color evolution}\label{sec:non_abelian}
In this section we assume that the initial quark probe is in the fundamental $SU(3)$ representation. As a consequence, the $H_A$ component of the Hamiltonian now has a non-trivial color structure, i.e. $A\cdot T=A^a T^a=\frac{1}{2}A^a \lambda^a$, where $\lambda^a$ denotes the eight Gell-Mann matrices. To deal with this modification, we further split the time evolution operator to take the form $U=U_K\cdot U_{A^1} \cdot U_{A^2}\cdots U_{A^8}$. Additionally, we must track the color of the quark as it evolves. To do that, we add a new register with two qubits, which stores the color state of the quark. In particular we use the following map between the logical and physical states: $\ket{0,0}\equiv \ket{\rm red}=\ket{R}$, $\ket{0,1}\equiv \ket{\rm green}=\ket{G}$, $\ket{1,0}\equiv \ket{\rm blue}=\ket{B}$ and $\ket{1,1}\equiv \ket{W}$, with the latter state not being physical and therefore absent from any calculation.

\begin{figure}[h!]
    \centering
    \includegraphics[width=0.8\textwidth]{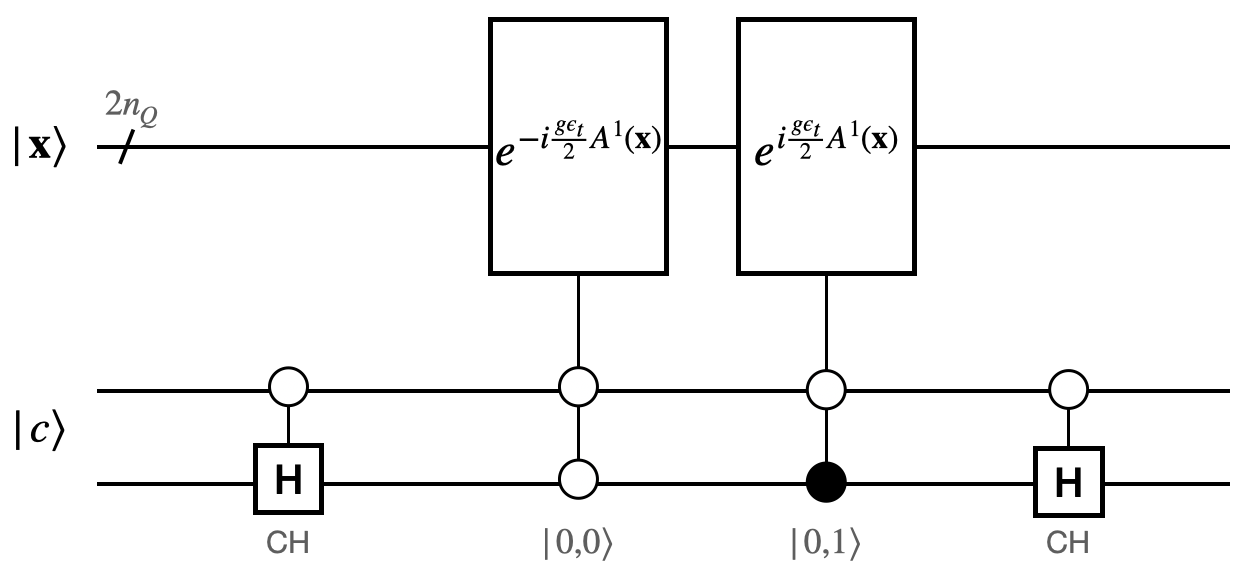}
    \caption{Implementation of the (infinitesimal) time evolution operator generated by $H_{A^1}$.}
    \label{fig:colors}
\end{figure}

We now detail how to implement $H_{A^1}$, with the other values of $a$ following analogous implementations. The first Gell-Mann matrix is given by 
\begin{equation}
\lambda^1=\begin{pmatrix}
0 & 1 & 0 \\
1 & 0 & 0 \\
0 & 0 & 0
\end{pmatrix}   \to
\begin{pmatrix}
0 & 1 & 0&0 \\
1 & 0 & 0 &0\\
0 & 0 & 0&0 \\
0 & 0 & 0&0 
\end{pmatrix} \equiv \Tilde{\lambda}^1 \, ,
\end{equation}
where in the second step we have embedded $\lambda^1$ into the two qubit Hilbert space. The action of $\Tilde{\lambda}^1$ is thus to color rotate the quark state between the $\ket{R}$ and $\ket{G}$ states, which can lead to a non-trivial evolution in color space. One can diagonalize the above matrix using a control Hadamard gate $CH$
\begin{equation}
CH=\begin{pmatrix}
1/\sqrt{2} & 1/\sqrt{2} & 0&0 \\
1/\sqrt{2} & -1/\sqrt{2} & 0 &0\\
0 & 0 &1&0 \\
0 & 0 & 0&1 
\end{pmatrix}   \, , 
\end{equation}
such that we can write $H_{A^1}$, in $k_t^{\rm th}$ time interval, in terms of a diagonal operator (here we drop all spacetime dependence for readability)
\begin{equation}\label{eq:colors}
e^{-\frac{ig\varepsilon_t}{2} A^1\otimes \tilde{\lambda}^1 }=(1\otimes  CH) e^{-\frac{ig\varepsilon_t}{2}  A^1\otimes \tilde{\sigma}^Z}  (1\otimes  CH)  \, .
\end{equation}
Here we made use of the extended Pauli operator $\tilde{\sigma}^Z={\rm diag}(1,-1,0,0)$\footnote{To be more precise, this definition takes $\tilde{\sigma}^Z$ to be non-unitary, unlike $\sigma^Z$. This is done, in order to ensure that only the $\ket{R}$ and $\ket{G}$ states transform non-trivially.}. Finally, to compute the exponential of the tensor product we notice that 
\begin{equation}
\begin{split}
e^{-i\frac{g\varepsilon_t}{2}A^1\otimes \tilde{\sigma}^Z}\ket{\x}\otimes \ket{c}&=\sum_n \frac{(-ig\varepsilon_t )^n}{2^n n!}  (A^1(\X)\tilde{\sigma}^Z)^n  \ket{\x} \ket{c} =\ket{\x}\sum_n \frac{(-ig\varepsilon_t  A^1(\x))^n}{2^n n!}   (\tilde{\sigma}^Z)^n   \ket{c} \, ,
\end{split}    
\end{equation}
where $\ket{c}$ denotes the two qubits register storing the state of the quark in color space. From the previous equation it is easy to observe that only $\ket{0,0}$ and $\ket{0,1}$ states result in a phase, the former with a $-i$ prefactor and the latter with a $+i$. Notice however, that due to the application of the diagonalizing gate $1\otimes  CH$, the evolution in the physical RGBW basis is off-diagonal. The implementation of Eq.~\ref{eq:colors} is given in Fig.~\ref{fig:colors}.

Clearly this strategy is only possible as long as the quark is in a small color representation -- in the previous example, the color degrees of freedom were treated by adding only two extra qubits and doubling the number of time evolution operators $U_{A^a}$, for each $a$. 

Another important consequence of including non-trivial color evolution is the fact that the final and initial state are differential in color. Therefore, when preparing the state one has to set colors either according to some initial state prescription or in an equitative way. Consequently, in the measurement protocol the output must be color averaged, which can be performed classically\footnote{This is not necessary if the qubits storing the color information are not measured.}.

\section{Numerical estimates for the circuit parameters}\label{sec:numerical_estimates}

In the previous sections we gave a conceptual outline on how to quantum compute the single particle momentum broadening distribution and from it extract meaningful physical information. Here we give a rough estimate on the typical values for the circuit parameters based on estimates for the typical physical scales obtained from jet quenching/saturation physics phenomenology.

Let us first estimate the lattice spacing $a_s$ and the number of qubits necessary per dimension $n_Q$. For that we recall that, when traversing a dense medium of length $L$, the quark will acquire an average transverse momentum of the order of the saturation scale, $\langle \p^2\rangle \sim \hat{q}L\equiv Q_s^2$. We are interested in length scales $L$ of the order of the nuclear radius of heavy elements, like Pb or Au, which we take to be $L\sim \mathcal{O}(10 \, \rm fm)=\mathcal{O}(50 \, \rm GeV^{-1})$. The value of the jet quenching parameters $\hat{q}$ varies drastically between different experimental set-ups, due to the different energy scales being explored. To bridge RHIC, LHC and EIC experimental conditions, we assume that $\mathcal{O}(0.1\, \rm GeV^2fm^{-1})\leq\hat{q}\leq\mathcal{O}(10\, \rm GeV^2fm^{-1})$~\cite{Feal:2019xfl,Liu:2018trl,Barata:2020jtq}. The saturation scale $Q_s^2$ is then approximately bounded by $Q_s^2\sim\mathcal{O}(1-100 \, {\rm GeV}^2 )$.

Setting the ultraviolet momentum cutoff induced by the digitization $\p_{\rm max.}$ to be much larger than the saturation scale $Q_s$, we obtain
\begin{equation}
|\p_{\rm max.}|\approx \frac{2\pi}{a_s}\gg \mathcal{O}(1-10 \, {\rm GeV}) \, ,
\end{equation}
thus 
\begin{equation}
 a_s\ll \mathcal{O}(1- 10\, {\rm GeV^{-1}})=   \mathcal{O}(0.1- 1\, {\rm fm})   \, .
\end{equation}
 Conversely, we require that the momentum space discretization is neither too coarse nor too fine. A simple way to ensure this is to impose
 \begin{equation}
\mu < a_d< Q_s  \sim  \frac{\mu}{Q_s} < \frac{1}{N_s}<1   \, ,
 \end{equation}
 with $\mu$ an infrared (model dependent) regulator, related to the medium Debye mass; typically $\mu~\sim \mathcal{O}(0.1-1\, \rm GeV)$~\cite{Barata:2020jtq,GW,HTL} and we used the previous estimates to reduce the problem to the ratio between the soft and hard scales at play. Recalling that $N_s=2^{n_Q}$, we obtain
 \begin{equation}
1< N_s< 100 \iff 0< n_Q< 7\, .     
 \end{equation}
Thus, one roughly needs $\mathcal{O}(2^7=128)$ states per dimension to adequately discretize the problem. In practice this number will have to be larger since the correct energy ratio should be $\mu/|\p_{\rm max.}|$, which here we took $|\p_{\rm max.}|=Q_s$. This is a rather rough lower bound, and larger values should be considered such that the peak of the broadening distribution is well captured. Even so, one would expect that (roughly) $n_Q< 20$, which means that $N_s<\mathcal{O}(10^6)$. This allows us to argue that the classical operations detailed in the previous sections needed to implement $U_A$ can be performed in a classical computer.

Let us now consider the longitudinal scales entering the problem and estimate the number of time steps $N_t$. We recall that in the multiple soft scattering approximation, one usually requires that the mean free path of the quark $\lambda$ is much larger than the typical correlation length in the medium $1/\mu$. This ensures that spatially delocalized scattering centers are not color correlated. On the other hand we also have that in order for a scattering to occur $\lambda\leq L$, leading to
\begin{equation}
1\geq \frac{\lambda}{L}\gg \frac{1}{\mu L}\, .  
\end{equation}
It is also typical to define the opacity of the medium as $\chi_{\rm med.}\equiv L/\lambda$~\cite{Gyulassy:2000er,Wiedemann:2000za}, corresponding to the expected number of in-medium scatterings. Therefore, it is natural to identify $\chi_{\rm med.}\sim N_t=L^\prime/\varepsilon_t$. We can then write 
\begin{equation}
1\leq N_t\ll \mu L \implies 1\leq N_t\ll \mathcal{O}(100)\, .  
\end{equation}

The remaining circuit parameter that directly depends on the physics one wants to explore is $m$, the number of field configurations to be generated. As alluded above, the numerical value for $m$ intrinsically depends on the model/prescription for the gauge field and its fluctuations, and therefore it is tied to the its physical origin. As such, and since in our treatment we have avoided discussing details of the background, we leave an estimation of this parameter for future work where a model for $\cA$ is chosen.

\section{Conclusions and Outlook}
\label{sec:Conclusion}

In this paper we have outlined a quantum simulation algorithm to extract the jet quenching parameter $\hat{q}$. Our approach is a hybrid one, with the in-medium jet evolution being quantum simulated, while the background field is treated as an external stochastic parameter, given as an input to the algorithm. The connection to \textit{standard} jet quenching language is immediate, unlike more recent efforts which rely on open quantum system formulation of jet quenching~\cite{deJong:2020tvx}, still not fully developed (see however~\cite{Vaidya:2020cyi,Vaidya:2021mjl}). 

The overall algorithm requires $2n_Q+l$ qubits (assuming one can re-use ancillas) and $\mathcal{O}(N_t\times {\rm polylog} N_s)$ basic gate operations. However, there is an underlying classical cost coming from the $m\times N_t\times N_s^2$ evaluations of the gauge field. This is the major drawback of our strategy, since it is not guaranteed that the classical evaluations of $\cA$ can be performed efficiently. Additionally, there is an overall additional polynomial cost in the measurement section, if one decides to scan several values of $\alpha$. For an actual implementation in a NISQ device~\cite{preskill2018quantum}, these constraints should not be limiting and we hope in the future more efficient algorithms can also be found. Nonetheless, we expect that our method can not outperform current classical approaches. 

In future work, we plan to implement our strategy in one spatial dimension (assuming azimuthal symmetry), bench-marking to known results for $\hat{q}$ from jet quenching phenomenology. This would allow a better understanding of the merits of our quantum approach, compared to known classical methods.

Going beyond $\alpha_s^0$ effects is of course of extreme relevance and the main motivation of our work. Indeed, since broadening is a classical effect, there is little advantage in applying quantum computing techniques to study it. However, a natural but non-trivial next step would be to include parton branching into the evolution operator. This is a purely quantum effect. If one is able to quantum simulate such a process efficiently, then interference contributions, inaccessible to classical Monte Carlo codes, can be exactly taken into account. The major obstacle to overcome is the fact that particle number is no longer conserved, and thus a new formulation of the problem is needed. Nonetheless, since broadening is a key element of in-medium propagation, the present algorithm provides a first step in this direction.

\begin{acknowledgements}
 JB is grateful to Niklas Mueller, Raju Venugopalan and Andrey Tarasov for helpful discussions on QIS and to Yacine Mehtar-Tani for pointing out quantum simulating broadening could be relevant. We also are grateful to Nestor Armesto for useful comments. JB is supported by a fellowship from ``la Caixa" Foundation (ID 100010434). The fellowship code is LCF/BQ/ DI18/11660057. This work has received financial support from European Union’s Horizon 2020 research and innovation program under the grant agreement No. 82409; from Xunta de Galicia (Centro singular de investigación de Galicia accreditation 2019-2022); from the European Union ERDF; from the Spanish Research State Agency by “María de Maeztu” Units of Excellence program MDM-2016-0692 and project FPA2017-83814-P and from the European Research Council project ERC-2018-ADG-835105 YoctoLHC
\end{acknowledgements}


\bibliographystyle{elsarticle-num}
\bibliography{Lib.bib}   

\appendix

\section{Discretization and encoding details}
\label{app:discretization}
In this appendix we give the details on the discretization of the quantum mechanical system considered in the main text and the map to the qubits available in the quantum computer. The discussion in this appendix is quite standard and can be found in quantum mechanics/computing textbooks~\cite{Sakurai:2011zz,NielsenChuang}.

We discretize space using a two dimensional lattice, with lattice spacing $a_s$ and $N_s$ lattice sites per dimension, with the spatial cut-off (per dimension) given by $a_s(N_s-1)$. We write the position ket $\ket{\x}=\ket{a_s\n}$\footnote{Notice that $\ket{\x}$ has the same mass dimension as $\x^{-1}$.}, with $\n$ a two dimensional integer vector. Conversely, the momentum induced lattice has a spacing $a_d=\frac{2\pi}{N_sa_s}$ (with an associated dimensionless integer vector $\q$) and the two bases are related by a Fourier transform
\begin{equation}
\begin{split}
|\p\rangle&= \int_\x e^{-i \p\cdot \x} |\x\rangle \to a_s^2 \sum_\n e^{-2\pi i\frac{ \q\cdot \n}{N_s} } |\n a_s\rangle \, ,
\end{split}    
\end{equation}
\begin{equation}
\begin{split}
|\x\rangle&= \int_\p e^{i \p\cdot \x} |\p\rangle \to \frac{a_d^2}{(2\pi)^2}\sum_\q e^{2\pi i \frac{\q\cdot \n }{N_s}} |\q a_d\rangle \, ,
\end{split}    
\end{equation}
where $\int_\x=\int d^2\x$ and $\int_\p=\int (2\pi)^{-2}d^2\p$ and we provide the discretized version of the Fourier integrals. Using that
\begin{equation}
\langle \x|\p\rangle=e^{-i\p\cdot \x}    \, \to \, e^{-2\pi i\frac{\n\cdot\q}{N_s}} \, ,
\end{equation}
one can show that 
\begin{equation}
\langle \x|\y\rangle=\delta^{(2)}(\x-\y)=\frac{\delta_{\n,\m}}{a_s^2}\, , 
\end{equation}
\begin{equation}
\langle \p|\k\rangle=(2\pi)^2\delta^{(2)}(\k-\p)=(2\pi)^2\frac{\delta_{\q_\k,\q_\p} }{a_d^2} \, ,
\end{equation}
where we used the closure identity 
\begin{equation}
\sum_{\n} e^{2 \pi i \frac{\n \cdot \q}{N_s}}=N_s^2 \delta_{\q,0}    \, .
\end{equation}
We define the dimensionless basis states
\begin{equation}
\begin{split}
\ket{\n}= a_s\ket{\x} \,, \quad \ket{\q}=\frac{a_d}{2\pi}\ket{\p} \, ,
\end{split}    
\end{equation}
which satisfy $\langle \n|\m\rangle=\delta_{\n,\m}$, $\langle \q_\p|\q_\k\rangle=\delta_{\q_\p,\q_\k}$ and $\langle \n|\q\rangle=N_s^{-1}\exp(-2\pi iN_s^{-1}\n \cdot \q)$. The Fourier transforms in this normalization take the form 
\begin{equation}\label{eq:QFT1}
\ket{\n}=\frac{1}{\sqrt{N_s^2}}\sum_{\q} e^{2\pi i\frac{ \q \cdot \n}{N_s}}   \ket{\q}    \, ,
\end{equation}
\begin{equation}\label{eq:QFT2}
\ket{\q}=\frac{1}{\sqrt{N_s^2}}\sum_{\n}e^{-2\pi i \frac{\q \cdot \n}{N_s}}  \ket{\n}    \, .
\end{equation}
It is also natural to introduce the operators
$\P=\p/a_d$ and $\X=\x/a_s$, satisfying $\hat{\X}\ket{\n}=\n\ket{\n}$ and $\hat{\P}\ket{\q}=\q\ket{\q}$. Inserting this operator definitions into Eq.~\ref{eq:Hamiltonian}, one can extract the dimensionless Hamiltonian $H=a_s \cH$, given in Eq.~\ref{eq:dimless_H}. 

The map to the $1/2$-spin registers in the quantum computer is achieved by decomposing each component of the vector $\n=(n_1,n_2)$ in the binary basis, e.g.
\begin{equation}
n_1=\sum_{i=0}^{2^{n_Q}-1} n_1^{(i)}2^i \, ,     
\end{equation}
where $n_1^{(i)}\in\{0,1\}$ and we assume that there are $n_Q$ qubits available, such that $2^{n_Q}=N_s$ is total number of possible states. If $n_1^{(i)}=0$ then we associate a qubit in the state $\ket{\ua}=\ket{0}$ to it; conversely if $n_1^{(i)}=1$ we assign $\ket{\da}=\ket{1}$. Then, for example, the ket state $\ket{\n}=\ket{3,3}$ with $n_Q=2$ is given by two registers storing the overall state $\ket{1,1}\otimes \ket{1,1}$. Following Eqs.~\ref{eq:QFT1} and \ref{eq:QFT2}, the transformation between the momentum and position basis is achieved by applying a standard quantum Fourier transform (qFT)~\cite{NielsenChuang}. 

Finally, in this appendix and in the main text we have restricted ourselves to considering lattices over positive integer values of $\n$ and $\q$. In an actual implementation, one would have to consider signed values, since, in general, there is no condition that physically constrains the system to positive values. In principle, signed values can be dealt with by, for example, including an extra qubit that stores the sign of the state (similar to the encoding used in~\cite{Barata:2020jtq}) or using a two's complement encoding. This caveat requires one to modify circuits we detail in the main text to accommodate for the new encodings. In general, one should be able to do this without incurring in an exponential number of extra qubits or basic gate operations\footnote{See for example~\cite{Klco:2018zqz} for an example on how to restrict the qFT to the first Brillouin zone.}, nor does it lead to any new conceptual challenge that must be addressed. As such, we do not further discuss this issue in the paper and leave it for a future work where we tackle a detailed implementation of the algorithm.

\section{Time evolution details}
\label{app:Zalka}
In this appendix we detail the key steps to implement the time evolution operators $U_K$ and $U_A$. For convenience and clarity, and without loss of generality, we will discuss both cases in one spatial dimension.

The strategy considered to implement $U_K$ was first discussed in~\cite{Zalka_1998}. Starting from a state $\ket{\p}$ (with $\p$ now having a single component) one wants to generate the state $\exp(-i s_K \p^2)\ket{\p}$, with $s_K=\varepsilon_t/(2E)$ a pure real number which can be easily computed once all circuit parameters are fixed. This operation can be implemented by i) adding an ancilla register with $l$ qubits all in state $\ket{0}$ ii) assuming that a quantum black-box (quantum oracle) can be constructed that given $\ket{\p}$ outputs $\ket{F(\p)}=\ket{\p^2}$.

Regarding the first point, the value of $l$ solely depends on the numerical accuracy one wants to represent $\p^2$ in a binary basis, roughly $l\geq n_Q$. An efficient quantum oracle implementing the above operation can always be constructed as long as a classical analog exists; this is the case for the operation at hands. 

Given both these conditions are satisfied, we then perform the following set of operations
\begin{ceqn}
\begin{equation}
\ket{\p}\otimes \ket{0}^{\otimes l}\stackrel{\rm a_1}{\longrightarrow}  \ket{\p}\otimes \ket{F(\p)}
\stackrel{\rm a_2}{\longrightarrow} \exp(-is_K F(\p))\ket{\p}\otimes\ket{F(\p)}\stackrel{\rm a_3}{\longrightarrow}  \exp(-is_K F(\p))\ket{\p}\otimes\ket{0}^{\otimes l} \, .
\end{equation}
\end{ceqn}
Let us detail the above three steps. In a first step --$\rm a_1$-- one applies the quantum oracle, with input $\ket{\p}$ and stores the output $F(\p)$ in the ancilla register. In step $\rm a_2$ one performs a transformation of the form
\begin{equation}
 \ket{x}\to\exp(-i s_K x)\ket{x} \, ,  
\end{equation}
with $s_K$ a real number and $\ket{x}$ denotes the binary decomposition, with $l$ qubits, of an integer number. This exponentiation operation can always be performed by applying $l$ single qubit gates $R_j(\varepsilon)={\rm diag}(1, e^{-is_K 2^j})$, taking into account that $x$ can be decomposed as
\begin{equation}
x=\sum_{j=0}^l x_j 2^j  \, ,
\end{equation}
where $x_j\in\{0,1\}$. Acting on a single qubit the above operator has non-zero matrix elements $\bra{0}R_j(s_K)\ket{0}=1$ and $\bra{1}R_j(s_K)\ket{1}=\exp(-i s_K 2^j)$; clearly stringing together $l$ of such operators with increasing values of $j$
\begin{equation}
R(s_K)\equiv R_0(s_K) \otimes R_1(s_K) \otimes \cdots \otimes R_l(s_K)\, ,    
\end{equation}
results in a multi-qubit operator implementing the desired transformation, i.e. $R(s_K)\ket{x}=\exp(-i s_K x)\ket{x} $. 

The final step -- $\rm a_3$ -- consists in erasing the ancilla register back to the state $\ket{0}^{\otimes l}$, which can be achieved by applying the Hermitian conjugate circuit used in step $\rm a_1$. 

The implementation of the operator $U_A$ could be done following exactly the same strategy as just described. However, as mentioned in the main text, this would require having a way to construct efficient quantum oracles that, for each time $t$, given $\ket{\x}$ output $\ket{A(t,\x)}$. We expect that for most cases, this will be difficult to do. 

As an alternative we consider that one is handed a list of $N_t \times N_s^2$ values, describing the field values at all the relevant spacetime points. Then one can implement $U_A$ by stringing together $2n_Q$ single qubit gates $R_{\alpha,\beta}\equiv{\rm diag}(\exp(i\alpha),\exp(i\beta))$. Such gates can be written as the product of the exponential of Pauli gates and the $R_j$ gate. In one spatial dimension and for $n_Q=1$ one simply has for the $k_t^{\rm th}$ time slice that $\alpha_{k_t}=-g\epsilon_t A(k_t\cdot \varepsilon_t,\0)$ and $\beta_{k_t}=-g\varepsilon A(k_t\cdot \varepsilon_t,\mathbf{1})$, where the sub-index denotes the time slice and there are only two spatial lattice points ($\ket{\0}$ and $\ket{\mathbf{1}}$). If we now consider $n_Q=2$ but still a single spatial dimension, the respective time evolution operator would be obtained by
\begin{equation}
 R_{\alpha,\beta}\otimes R_{\sigma,\gamma}=\begin{pmatrix}
{\rm e}^{i(\alpha+\sigma)} & 0 & 0 & 0\\
0 & {\rm e}^{i(\alpha+\gamma)} & 0 & 0\\
0 & 0 & {\rm e}^{i(\beta+\sigma)} & 0\\
0 & 0 & 0 & {\rm e}^{i(\beta+\gamma)}\\
\end{pmatrix} \, ,    
\end{equation}
for each time slice. By solving the associated system of linear equations, one can map $\{\alpha,\beta,\sigma,\gamma\}$ to $\{A(\mathbf{x})\}$, which can be done offline for any $t$ in a classical computer.

\section{Relation between $\ket{\psi_L}$ and the single particle momentum distribution}
\label{app:broad}
In this appendix we relate $\ket{\psi_L}$ to the broadening distribution. 

The single particle broadening probability for observing a quark with momentum $\k$ due to interactions with the medium for a time $L$ is given by~\cite{Blaizot:2012fh,Blaizot:2015lma,CasalderreySolana:2007zz}
\begin{equation}
\cP(L,\k)=\frac{1}{N_c} \int_{\x,\y} \rme^{-i\k\cdot (\x-\y)} \Tr \langle \cW(L,\x) \cW^\dagger(L,\y) \rangle_{\rm M} \, ,  
\end{equation}
where $\cW(L,\x)$ is a Wilson line operator along the future pointing light-cone at a transverse position $\x$, which can be written in the gauge choice employed in the main text as
\begin{equation}
\cW(L,\x)=\cT\exp\left(ig\int_0^L dt \, \cA^-(t,\x)\cdot T\right) \, .
\end{equation}
The above medium average is usually performed by detailing the non-trivial correlators of the background field, in jet quenching typically the MV/Gaussian prescription. Using this further assumption, one can then write the broadening distribution in terms of a so called \textit{dipole cross-section}, which is typically constrained to recover the Coulomb form at short distances and to have a model dependent form in the infrared~\cite{GW,HTL}.

It is not difficult to check that, in the strict eikonal limit, where $H=H_A$, the circuit detailed in the main text mirrors the $\cP$ distribution. For clarity, we ignore the details in the implementation of the time evolution operator; additionally, we assume that the initial state is that of a quark with zero transverse momentum $\ket{\psi_0}=\ket{\p=\0}$. 

In this scenario the circuit simplifies significantly since all but an initial and a final qFT cancel out. Then the system state transforms as
\begin{ceqn}
\begin{equation}
\ket{\0}\stackrel{\rm qFT}{\longrightarrow}  \frac{1}{\sqrt{N_s^2}}\sum_\x \ket{\x} \stackrel{U_A}{\longrightarrow}  \frac{1}{\sqrt{N_s^2}}\sum_\x U_A(L,\x)\ket{\x}\stackrel{\rm qFT^\dagger}{\longrightarrow}\frac{1}{N_s^2}\sum_\q \left[\sum_\x U_A(L,\x) {\rm e}^{2\pi i \frac{\x \cdot \q}{N_s}}\right]\ket{\q} \, .
\end{equation}
\end{ceqn}
The probability of measuring the state $\ket{\k}$, $\cP_\k$, is simply given by 
\begin{equation}
\cP_\k=\frac{1}{(N_s^2)^2}\sum_{\x,\y}{\rm e}^{2 \pi i \frac{\k(\x-\y)}{N_s}}  U_A^\dagger(L,\y)U_A(L,\x) \, .
\end{equation}
Averaging over all field configurations and noticing that $\cW(\x)=U_A^\dagger(\x)$ we obtain
\begin{equation}
\cP_\k=\frac{1}{(N_s^2)^2}\sum_{\x,\y}{\rm e}^{2 \pi i \frac{\k(\x-\y)}{N_s}}  \langle \cW(L,\y)\cW^\dagger(L,\x)\rangle_{\rm M} \, ,
\end{equation}
which is just the discretized version of the single particle broadening distribution $\cP(L,\k)$, as expected (ignoring the color average, which can be performed as detailed in section~\ref{sec:non_abelian}). Also, since $\cP$ is a probability $\int_\k \cP(L,\k)=1$, which is trivially true in the discrete version.

Notice that in our strategy, one does not need to explicitly provide a prescription for the field correlators. Of course, these are embedded in the generated field configurations and are taken into account (non-perturbatively) by the time evolution operator.

\section{Measurement details}\label{app:meas}
In this appendix we provide some details on the measurement protocol outlined in the main text.

Taking the initial ancilla state to be $\ket{0}$, the measurement protocol performs 
\begin{equation}
\begin{split}
\ket{0}\ket{\psi_L}
\stackrel{V \, H}{\longrightarrow}\frac{1}{\sqrt{2}}(\ket{0}\ket{\psi_L}+\ket{1}V\ket{\psi_L})\stackrel{H}{\longrightarrow}\frac{1}{2}\left[ (1+V)\ket{0}\ket{\psi_L}+(1-V)\ket{1}\ket{\psi_L}\right] \, .
\end{split}
\end{equation}
Then the expectation value for the random variable $\chi$ reads
\begin{equation}
\begin{split}
\langle \chi \rangle_{\rm QM}=\frac{+1}{4}|\ket{\psi_L}+V\ket{\psi_L}|^2+\frac{(-1)}{4}|\ket{\psi_L}-V\ket{\psi_L}|^2=\frac{1}{2}\langle V+V^\dagger\rangle_{\rm QM} \, ,
\end{split}
\end{equation}
which is equivalent to the expression in the main text. 

The case where the initial ancilla state is $\ket{\gamma}\equiv1/\sqrt{2}(\ket{0}+i\ket{1})$, which can be easily generated from the pure state $\ket{0}$, reads
\begin{equation}
\begin{split}
\ket{\gamma}\ket{\psi_L}&\stackrel{VH}{\longrightarrow}\frac{1}{2}((1+i)\ket{0}\ket{\psi_L}+(1-i)\ket{1}V\ket{\psi_L}) \\&\stackrel{H}{\longrightarrow}\frac{1}{\sqrt{8}}[ ((1+i)+(1-i)V)\ket{0}\ket{\psi_L}+((1+i)-(1-i)V)\ket{1}\ket{\psi_L}] \, . 
\end{split}
\end{equation}
Then the expectation value for $\chi$ reads 
\begin{equation}
\langle \chi\rangle_{\rm QM}=\frac{i}{2} \langle V^\dagger-V\rangle_{\rm QM} \, ,  
\end{equation}
as indicated in the main text.

\end{document}